\documentclass{article}
\PassOptionsToPackage{numbers,sort&compress}{natbib}

\usepackage[preprint]{neurips_2026}

\usepackage[utf8]{inputenc}
\usepackage[T1]{fontenc}
\usepackage[dvipsnames]{xcolor}
\usepackage{graphicx}
\usepackage{flafter}
\usepackage{hyperref}
\usepackage{url}
\usepackage{booktabs}
\usepackage{amsmath}
\usepackage{amsfonts}
\usepackage{nicefrac}
\usepackage{microtype}
\usepackage{algorithm}
\usepackage{algpseudocode}
\usepackage{placeins}
\usepackage{multirow}
\usepackage{booktabs}
\usepackage{tabularx}
\usepackage[most]{tcolorbox}
\newtcblisting{PromptBox}[1]{
  enhanced,
  breakable,
  listing only,
  colback=green!3!white,
  colframe=ForestGreen!70!black,
  coltitle=black,
  fonttitle=\bfseries,
  title={#1},
  boxed title style={colback=green!15!white,colframe=ForestGreen!70!black},
  attach boxed title to top left={xshift=1mm,yshift=-2mm},
  boxrule=0.6pt,
  arc=1mm,
  left=1mm,
  right=1mm,
  top=1mm,
  bottom=1mm,
  before skip=8pt,
  after skip=8pt,
  listing options={
    basicstyle=\ttfamily\small,
    breaklines=true,
    breakatwhitespace=false,
    columns=fullflexible,
    keepspaces=true,
    showstringspaces=false
  }
}
\newtcolorbox{TakeawayBox}[1][Takeaway]{
  enhanced,
  breakable,
  colback=green!3!white,
  colframe=ForestGreen!70!black,
  coltitle=black,
  fonttitle=\bfseries,
  title={#1},
  boxed title style={colback=green!15!white,colframe=ForestGreen!70!black},
  attach boxed title to top left={xshift=1mm,yshift=-2mm},
  boxrule=0.6pt,
  arc=1mm,
  left=1.5mm,
  right=1.5mm,
  top=1mm,
  bottom=1mm,
  before skip=8pt,
  after skip=8pt
}

\title{BootstrapAgent: Distilling Repository Setup into Reusable Agent Knowledge}
\author{%
  Sihan Fu$^{1,*}$ \quad
  Oucheng Liu$^{2,*}$ \quad
  Shiyuan Wang$^{3}$ \quad
  Jin Shi$^{1}$ \quad
  Chengkun Wei$^{1, \dagger}$ \\
  $^{1}$Zhejiang University \quad
  $^{2}$The Australian National University \\
  $^{3}$Institute of Information Engineering, Chinese Academy of Sciences \\
  \texttt{\{fusihan,shijin,weichengkun\}@zju.edu.cn} \quad
\\
\texttt{oucheng.liu@anu.edu.au} \quad
  \texttt{wangshiyuan@iie.ac.cn}\\
  \small {$^{*}$Equal contribution. $^{\dagger}$Corresponding author.}
}

\begin{document}

\maketitle

\begin{abstract}
Code agents increasingly help developers work with unfamiliar repositories, but every such task depends on a costly prerequisite: bootstrapping the repository into a usable development state. This process requires substantial trial-and-error exploration, yet the resulting knowledge---resolved dependencies, repair strategies---stays trapped in a single conversation, unavailable to future agents. We therefore formulate repository bootstrapping as a reusable startup knowledge problem and introduce \textbf{BootstrapAgent}, a multi-agent framework that distills the heuristics discovered during bootstrap exploration into a persistent, verifiable, agent-consumable \texttt{.bootstrap} contract. Through evidence extraction, structured planning, deterministic Docker-based verification, and trace-driven repair, BootstrapAgent generates a contract covering environment setup, diagnostic checks, minimal verification, and accumulated repair knowledge. We further propose \textit{warm repair with clean replay} to accelerate iterative debugging without sacrificing cold-start reproducibility, and a \textit{delta repair with sanity check} to prevent reward hacking. Experiments on three benchmarks show that BootstrapAgent achieves a 92.9\% success rate, outperforming the baseline by over 10\% while reducing downstream agent token usage by 25.9\% and build time by 22.3\%. Our code is available at \url{https://github.com/Vossera/BootstrapAgent}. 
\end{abstract}

\section{Introduction}
\label{sec:introduction}
\begin{figure}[t]
\centering
\includegraphics[width=0.9\linewidth]{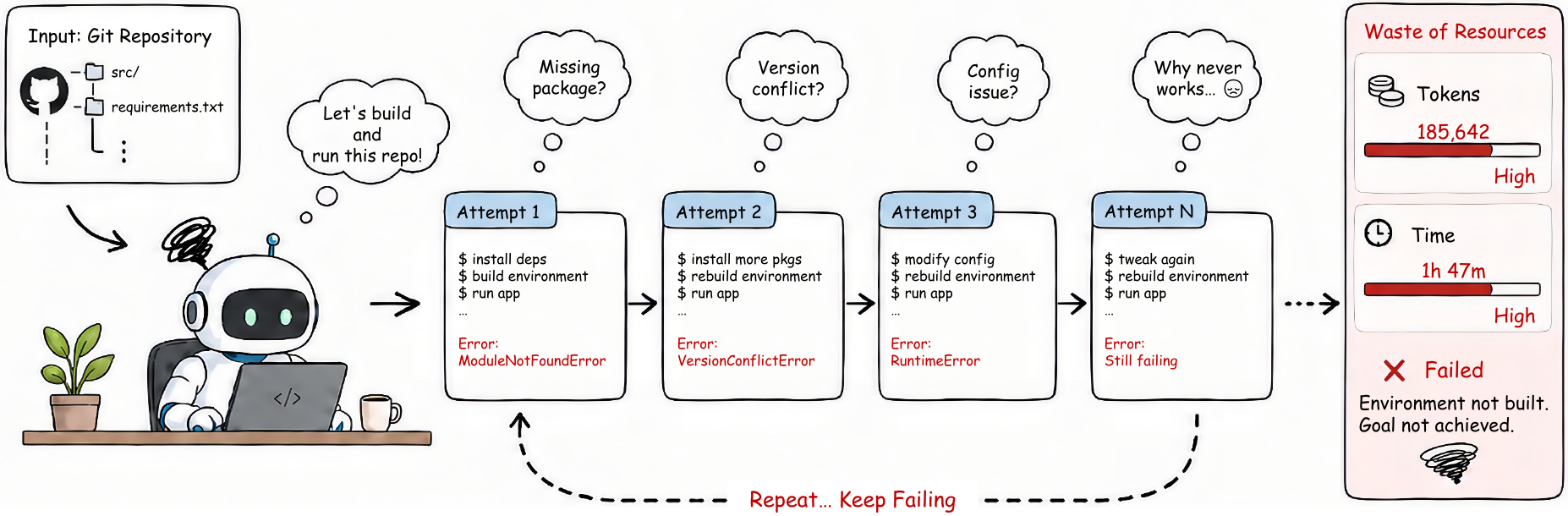}
\caption{The process of deploying a repo by Code Agent.}
\label{fig:intro1}
\end{figure}
Real-world development rarely starts from scratch but builds on existing repositories~\citep{para1-1,para1-2}, whether to extend features, reproduce experiments, migrate components, or build prototypes. Before any such task can begin, the target repository must first be brought into a usable development state. However, the path from a freshly cloned repository to a working environment is often unclear~\citep{kovrigin2025piper}. Existing repository artifacts provide useful clues, but these clues are typically incomplete~\citep{hu2025repo2run, heragent, installmatic}: README files may describe package usage rather than the development setup; CI (Continuous Integration) workflows may depend on secrets, caches, service containers, specialized hardware, or long-running matrix jobs that are difficult to reproduce locally; and package metadata rarely specifies the minimal command that can confirm a local checkout is actually usable.


Although code agents, with their emerging capabilities in repository understanding and code execution, are rapidly lowering the barrier for developers to work with unfamiliar codebases, repository bootstrapping remains a particularly costly bottleneck for these agents~\citep{codeagent-de-1, codeagent-de-2, codeagent-de-3, codeagent-de-4, sweagent1}. As shown in Figure ~\ref{fig:intro1}, a typical agent must read the README, inspect package metadata and CI workflows, infer setup commands, execute them, observe failures, repair the environment, and retry~\citep{test, guo2025swe, para2-1-java}. This loop consumes substantial tokens and time. More critically, the knowledge produced during this process is ephemeral: resolved dependencies, working-directory corrections, and successful repair strategies remain trapped in a single-round conversation, unavailable to the next agent or developer working on the same repository. Multiple agents may therefore repeatedly pay the same bootstrap cost. Recent agent-facing files such as \texttt{AGENTS.md}~\citep{agentmd} and \texttt{CLAUDE.md}~\citep{anthropic_claude_memory_2026} reduce downstream development token cost by encoding repository-specific coding guidance, but they do not address the earlier question of how an agent should reliably prepare, diagnose, and verify a repository itself.

Our key observation is that the outcome of bootstrap exploration should not vanish with the conversation, but should instead be persisted as reusable startup knowledge. To this end, we propose \textbf{BootstrapAgent}, a multi-agent framework that distills bootstrap exploration into a persistent, verifiable, agent-consumable \texttt{.bootstrap} contract. BootstrapAgent begins with a discovery agent that scans repository structure, README files, CI workflows, lockfiles, package metadata, and build configuration files to extract bootstrap-relevant evidence. A planning agent then converts this evidence into a structured bootstrap plan containing setup commands, diagnostic checks, and verification commands. Based on this plan, a contract generator agent produces a \texttt{.bootstrap} contract. The contract is then replayed stage by stage by a deterministic verifier in a clean Docker ~\citep{docker1} environment. If replay fails, the generator agent uses structured execution traces to apply feedback-driven minimal revisions to \texttt{.bootstrap}, and verification is re-run. This heuristic repair-and-verify loop continues until the \texttt{.bootstrap} contract passes clean replay or a stopping condition is reached. Two key design choices underpin this pipeline: \textit{warm repair with clean replay} allows fast iterative repair within the same container to reduce cost, but requires the final contract to pass cold-start verification in a fresh environment to ensure reproducibility; and a \textit{delta repair with sanity checks} constrains each repair round to evidence-supported updates over the existing file, rather than unconstrained regeneration. This prevents agents from achieving superficial success by weakening the verification command.

We evaluate BootstrapAgent on 212 repositories drawn from three benchmarks and compare it against other environment setup approaches and agent-only bootstrap baselines. BootstrapAgent achieves a 92.9\% clean-replay bootstrap success rate, outperforming HerAgent~\citep{heragent} by over 10\% (8.5 percentage points). When the generated contracts are reused by downstream Claude Code agents, token usage is reduced by 25.9\% and build time by 22.3\%. In summary, our contributions are as follows:

\begin{itemize}
    \item We formulate repository bootstrapping as a reusable startup knowledge problem for code agents, and introduce \texttt{.bootstrap} as an agent-consumable contract.

    \item We design BootstrapAgent framework that combines evidence extraction, structured planning, and deterministic verification with trace-driven repair. Warm repair, clean replay, and delta repair with sanity checks make it efficient, reproducible, and robust to reward hacking.

    \item We evaluate on 212 repositories and demonstrate strong bootstrap success rates, showing that the generated contracts significantly reduce repeated setup cost for downstream agents.
\end{itemize}

\section{Related Work}
\label{sec:related-work}
\subsection{Automated Repository Bootstrapping and Environment Setup.}
Prior work has studied repository execution from several angles, ranging from template- or rule-based Dockerfile generation~\cite{ye2021dockergen,horton2019dockerizeme} to learning-based environment construction~\cite{rosa2023automaticallygeneratingdockerfilesdeep}. More recent systems move beyond static generation by using execution feedback, sandbox testing, and repository context to infer dependencies and validation commands. RepoST~\cite{xie2025repost} and Repo2Run~\cite{hu2025repo2run} construct executable repository environments through context mining and sandboxed verification; Treefix~\cite{souza2025treefixenablingexecutiontree} repairs execution through prefix-tree search; and CXXCrafter~\cite{10.1145/3729386} targets the specialized build and dependency challenges of C/C++ repositories. Complementary benchmarks such as DI-BENCH~\cite{zhang2025dibenchbenchmarkinglargelanguage}, CSR-Bench~\cite{xiao-etal-2025-csr}, R2E~\cite{pmlr-v235-jain24c}, and R2E-Gym~\cite{jain2025r2egymproceduralenvironmentshybrid} further expose dependency inference, scientific software deployment, and executable-repository construction as important evaluation problems. These works establish that repository setup is a difficult and measurable task. However, they primarily treat environment construction as a one-time deployment or benchmark outcome. BootstrapAgent instead treats setup exploration as reusable agent knowledge: it persists the discovered setup, diagnostic checks, verification commands, provenance, and repair knowledge into a verifiable \texttt{.bootstrap} contract that can be cleanly replayed and reused by future agents.

\subsection{Agent Context Files and Repository-Facing Documentation}
\label{subsec:rw-context-files-documentation}
Coding agents ~\citep{ google_gemini_cli_2026, anthropic_claude_code_2026, openai_codex_2026} have demonstrated strong abilities on benchmarks such as SWE-bench~\citep{yang2025swebench, jimenez2024swebench} and follow-on repository-level evaluations~\citep{benchmark1, issueResolution_2.1, featureaddition_2.1, featureaddition2_2.1, test_2.1, codeperformance_2.1, security_2.1}. 
As coding agents have been adopted in unfamiliar repositories, agent developers and users increasingly rely on context files such as \texttt{AGENTS.md}~\citep{agentmd} and \texttt{CLAUDE.md}~\citep{anthropic_claude_memory_2026}. Persistent artifacts such as
 configuration files, documentation, and contextual instructions continuously shape   agent behavior~\citep{context1, context2,context3,context4,context5}. However, prior work has not studied how repository-facing context files should specify
and maintain the bootstrap process itself, including dependency installation, and recovery from setup failures

\section{BootstrapAgent}
\label{sec:method}

\subsection{Problem Formulation}
\label{subsec:problem-formulation}

We study repository bootstrapping as the problem of converting an unfamiliar repository into a startup contract that can be reused by future agents. Let \(R\) be a repository specified by a URL or a local path, and let \(E_R\) denote the observable bootstrap evidence extracted from \(R\), including documentation, package metadata, lockfiles, build files, scripts, project layout, and CI workflows. The bootstrap configuration system must synthesize a contract \(C\) from \(E_R\), without assuming access to maintainer knowledge or host-specific state. A \texttt{.bootstrap} contract is a repository-local artifact:
\[
C = (I, D, M, S, H),
\]
where \(I\) is an ordered sequence of setup commands, \(D\) is a sequence of read-only diagnostic checks, \(M\) is a mandatory minimal verification command, \(S\) is an optional strongest locally reproducible verification command derived from CI or build evidence, and \(H\) is the compressed repair knowledge preserved for future agents. Each command also records its provenance in \(E_R\). The contract exposes a simple execution protocol: run setup, run diagnostics, and then run verification.

Synthesis is mediated by a structured \textit{bootstrap plan} \(P\). The plan is generated from \(E_R\), fixes the setup phases and verification goals, and links each decision to its supporting evidence, without committing to concrete shell commands. An initial contract \(C_0\) is materialized from \(P\), after which the system refines it through a bounded feedback loop. At iteration \(t\), the contract \(C_t\) is executed by a verifier \(V\) in a fresh container with base environment \(B\), producing an execution trace:
\[
\tau_t = V(R, C_t;\, B),
\]
which contains stage outcomes, exit codes, output, and failure locations. Conditioned on the fixed plan, the repair step proposes a trace-driven delta over the current contract:
\[
C_{t+1} = \mathrm{Repair}(C_t,\, P,\, \tau_t),
\]
while preserving commands and metadata that have already survived verification unless the trace provides direct evidence that they caused the failure. The plan \(P\) and evidence \(E_R\) constrain the feasible space of deltas: a sanity check rejects any revision that weakens an evidence-supported validation target. The loop runs under explicit budgets and stops once the contract is accepted or a budget is exhausted. The resulting contract is then frozen and consumed by downstream agents rather than further updated. In this sense, the iterative repair loop can be viewed as a heuristic system that operates under a bounded budget and terminates in a frozen, verified contract \citep{weng2026learning_beyond_gradients}. 

A contract is valid only if it passes \textit{clean replay}: given \(B\) and \(V\), the verifier runs \(C\) from a fresh container, with \(R\) mounted at a fixed path. Let \(Y_I\), \(Y_D\), and \(Y_M\) denote the stage outcomes of setup, diagnostics, and minimal verification, respectively. Then
\[
\mathrm{Valid}(R, C;\, B, V) =
\mathbb{I}\big[
Y_I = \textsc{pass}
\,\wedge\,
Y_D = \textsc{pass}
\,\wedge\,
Y_M = \textsc{pass}
\big],
\]
subject to safety checks that reject non-local assumptions, swallowed failures, and degenerate verification commands. The strongest command \(S\) is not required to pass---real CI may depend on secrets, services, special hardware, or long-running jobs---but serves as an advisory guardrail: repair must not weaken an identified project-relevant target merely to clear the minimal gate.

The goal is not to make one container pass once, but to produce a contract that is reproducible, auditable, and cheap to reuse:
\[
C^\star =
\arg\max_{C \in \mathcal{C}(E_R)}
\;
\mathrm{Valid}(R, C;\, B, V)
-
\lambda_1 \mathrm{Cost}_{\mathrm{gen}}(R, C)
-
\lambda_2 \mathrm{Cost}_{\mathrm{reuse}}(R, C),
\]
where \(\mathcal{C}(E_R)\) is the space of contracts reachable from \(E_R\) through plan-mediated synthesis and repair, \(\mathrm{Cost}_{\mathrm{gen}}\) is the cost of generating and repairing the contract, and \(\mathrm{Cost}_{\mathrm{reuse}}\) is the downstream cost for a fresh agent to reach the same verified startup state using the contract. BootstrapAgent operationalizes this objective with bounded repair budgets, deterministic Docker-based verification, warm repair for efficient search, and a final clean-replay requirement for acceptance.

\subsection{System Overview}
\label{subsec:system-overview}

\begin{figure}[t]
\centering
\includegraphics[width=\linewidth]{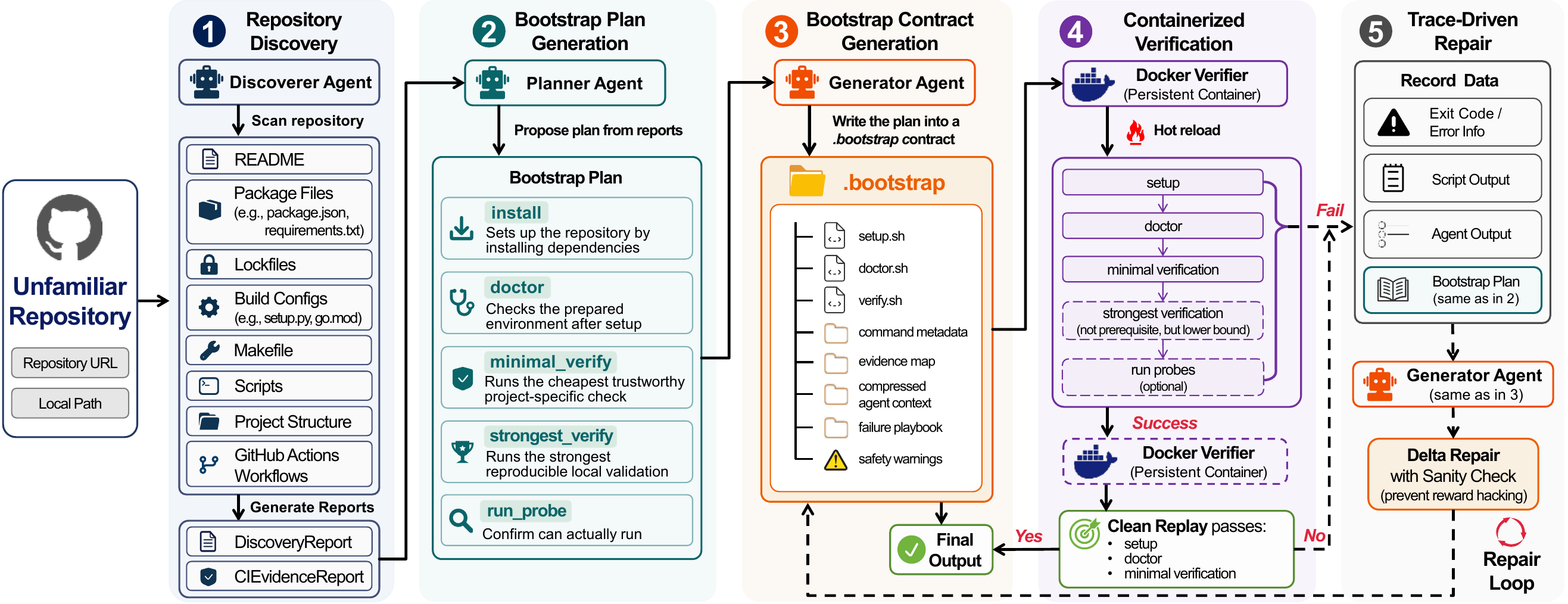}
\caption{Overview of BootstrapAgent.}
\label{Overview of BootstrapAgent}
\end{figure}

Given the above formulation, BootstrapAgent synthesizes a verified \texttt{.bootstrap} contract that specifies how to set up, inspect, and minimally verify a repository in a clean environment. As shown in Figure~\ref{Overview of BootstrapAgent}, this process proceeds through five stages: repository discovery, bootstrap plan generation, contract generation, containerized verification, and trace-driven repair. We describe each stage below.

\subsection{Repository Discovery}
\label{subsec:repo-discovery}

The first stage prepares the target workspace and collects repository evidence. A \textit{Discoverer Agent} explicitly scans files that are likely to affect bootstrapping, including \texttt{README} files, package metadata, lockfiles, build configuration files, Makefiles, scripts, project layout, and GitHub Actions workflows.

This stage produces two structured reports. \texttt{DiscoveryReport} records detected languages, package managers, important files, repository structure, and evidence snippets from relevant files. \texttt{CIEvidenceReport} records workflow files and local \texttt{run} commands that may serve as validation candidates. It also identifies non-local CI features, such as secrets, service containers, cloud services, or heavyweight external dependencies, and marks them as non-reproducible constraints that should not be selected for local verification.

\subsection{Bootstrap Plan Generation}
\label{subsec:bootstrap-plan}

Given the discovery and CI-derived reports, the \textit{Planner Agent}
produces a structured \texttt{BootstrapPlan}. It is a schema-constrained description of the intended bootstrap strategy. It records the expected setup phases, the verification goals, the evidence used to justify each decision, and the constraints that later agents must preserve. The plan acts as a persistent task anchor during repair. A repair agent may perform multiple rounds of reasoning and tool calls; without an explicit plan, it can drift toward fixing the most recent error while forgetting the higher-level bootstrap objective.

\subsection{Bootstrap Contract Generation}
\label{subsec:contract-generation}

This stage generates the \texttt{.bootstrap} directory, which contains \texttt{setup.sh}, \texttt{doctor.sh}, \texttt{verify.sh}, command metadata, an evidence map, agent context, a failure playbook, and safety warnings.  

Initially, no prior execution evidence exists, the \textit{Generator Agent} constructs a complete \texttt{.bootstrap} contract only based on the plan. After the initial contract is synthesized, the system enters a repair-and-verify loop. The scripts are executed inside a Docker container, producing an execution trace. The trace is then provided to the agent together with the plan to generate \texttt{.bootstrap}. This process repeats until the verification succeeds or the maximum number of repair rounds is reached.

\subsection{Containerized Verification}
\label{subsec:verification}

This stage executes the generated contract inside a Docker environment. It is necessary because repository bootstrap correctness cannot be reliably inferred from static evidence alone.  The repository is mounted at a fixed path, and the verifier runs setup, doctor, minimal verification, strongest verification, and optional run probes with stage-specific timeouts and logs. 

This staged execution has two competing requirements. During repair-and-verify loop, the system needs fast feedback so that small command-level changes can be tested without repeatedly paying the full setup cost. For final acceptance, however, the verifier must rule out accidental success caused by cached packages, generated files, or residual state from previous attempts. To satisfy both requirements, the \textit{Docker Verifier} uses a two-level correctness mechanism.

When in the repair loop, it uses \textbf{hot reload} in a Docker container to provide low-latency feedback for small structured plan deltas. This design is inspired by Spring Boot's hot-swapping workflow, where development tools monitor changes and support fast restarts or live reloads instead of repeatedly performing a full cold start~\cite{springboot_hotswapping}. The system then freezes the candidate contract and performs a \textbf{clean replay} from a fresh container. If the clean replay succeeds, the system outputs the final \texttt{.bootstrap} directory. If the clean replay fails, the system continues the repair loop. Thus, warm verification accelerates the search process, while clean replay preserves the semantic requirement that a bootstrap contract must work from an empty environment. 

\subsection{Trace-Driven Repair}
\label{subsec:tracedriven}

The \textit{Generator Agent} acts as the repair agent in this stage, but in a constrained repair mode. Instead of freely rewriting the entire contract, it proposes a structured delta over the existing \texttt{BootstrapPlan} and the current \texttt{.bootstrap} directory. A delta may insert a missing setup command, adjust command ordering, replace a failed command with an evidence-supported alternative, add diagnostic checks, or update timeout and fallback metadata. Commands and metadata that have already survived previous verifications are preserved unless the trace provides direct evidence that they caused the failure.

This delta-based design improves efficiency and stability across repair rounds. Because the agent only edits the failing or under-specified part of the contract, later rounds do not need to regenerate the entire \texttt{.bootstrap} directory. This reduces generation cost and limits the opportunity for hallucinated rewrites to degrade previously valid scripts, metadata, or evidence links. In other words, repair is treated as a localized update to an existing contract rather than as a new bootstrap attempt.

To further reduce reward hacking during multi-round repair, BootstrapAgent applies a sanity check after each proposed repair. The generator agent that repeatedly observes verifier failures may otherwise weaken the verification command to satisfy the acceptance criterion, for example by replacing a project-relevant build or test with a trivial version check. The sanity check compares the repaired contract against the original plan, the evidence map, and the strongest locally reproducible validation target. If the checks fail, the generator agent will regenerate a patch to .bootstrap. 

\section{Experiments}
\label{sec:experiments}
We use \texttt{DeepSeek-V4-Flash-2026-04-25} for all agentic planning and repair. The prompts used in our agent are shown in ~\ref{app:key-prompts}.

\subsection{Research Questions}
\label{subsec:research-questions}

Our evaluation is organized around the following research questions on three benchmarks. Detailed benchmark descriptions are given in Appendix~\ref{app:benchmark}.

\textbf{RQ1: Effectiveness.}
Can BootstrapAgent generate a \texttt{.bootstrap} contract that passes clean replay on real repositories?

\textbf{RQ2: Downstream bootstrap transfer.}
Does a generated \texttt{.bootstrap} contract reduce the cost for a fresh downstream agent to reach a verified startup state?


\textbf{RQ3: Ablation.}
Which design choices contribute to BootstrapAgent's performance?

\subsection{Effectiveness}
\label{subsec:effectiveness}
We measure the bootstrap success rate in a fresh container, where success requires \texttt{setup}, \texttt{doctor}, and \texttt{minimal\_verify} to pass. The detailed experiment settings are shown in Appendix~\ref{app:expset1}. Table~\ref{tab:effectiveness} reports the effectiveness of BootstrapAgent on the three public benchmarks. We compare
against HerAgent, the strongest reported baseline across these benchmarks. Across 212 repositories, BootstrapAgent improves over HerAgent by 8.5 percentage points, corresponding to a 10.1\% relative improvement. This indicates our multi-agent framework is effective for building most repositories.

\begin{figure*}[t]
	\centering
	\begin{minipage}[t]{0.32\linewidth}
		\centering
		\includegraphics[scale=0.30]{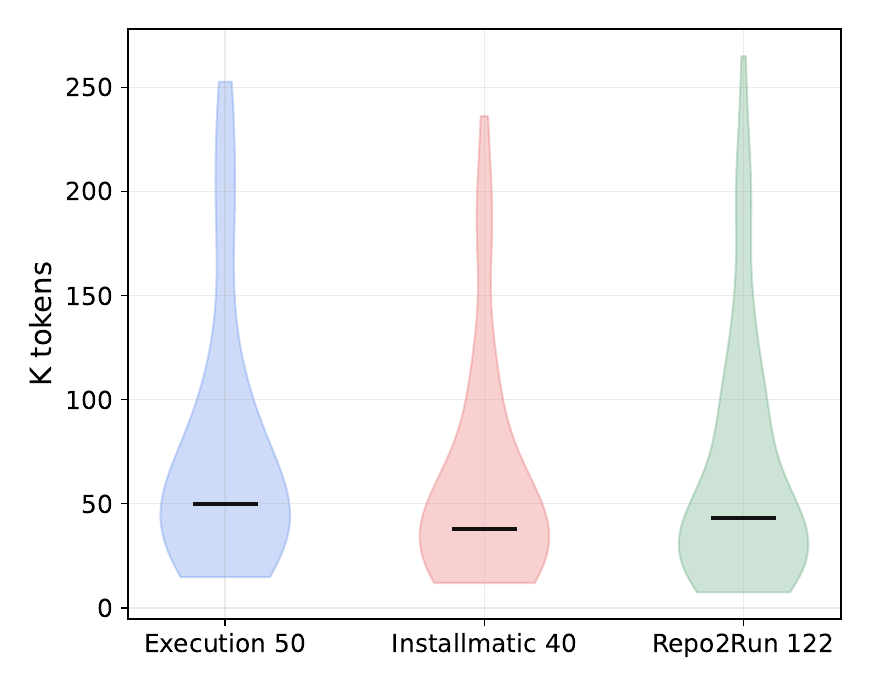}
        \caption{Token usage of BootstrapAgent.}
		\label{fig:tokens}
	\end{minipage}%
    	\hfill
	\begin{minipage}[t]{0.32\linewidth}
		\centering
        \includegraphics[scale=0.30]{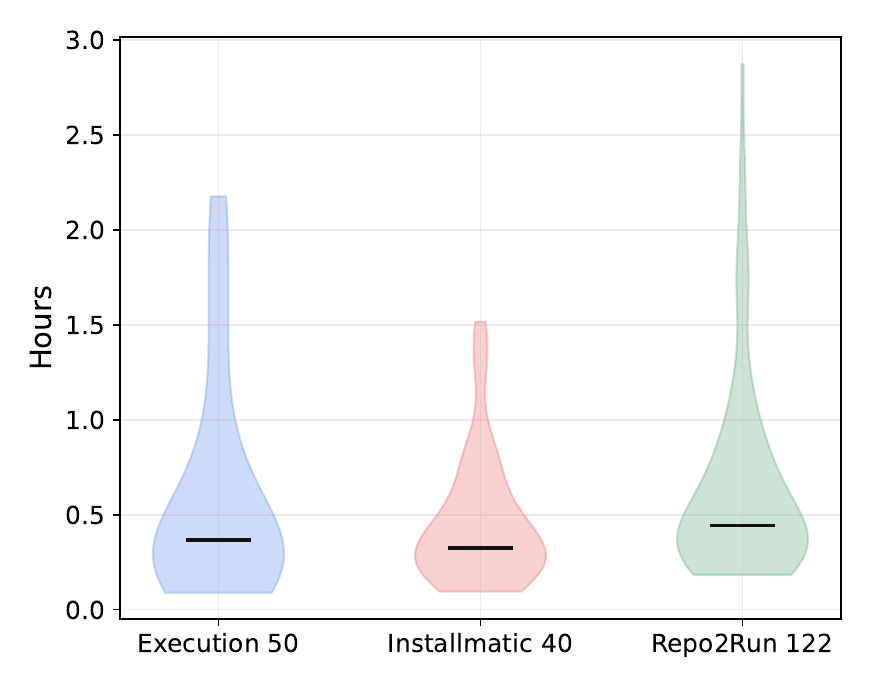}
        \caption{Time cost of BootstrapAgent.}        
        \label{fig:wallclock}
	\end{minipage}%
	\hfill
	\begin{minipage}[t]{0.32\linewidth}
		\centering
        \includegraphics[scale=0.30]{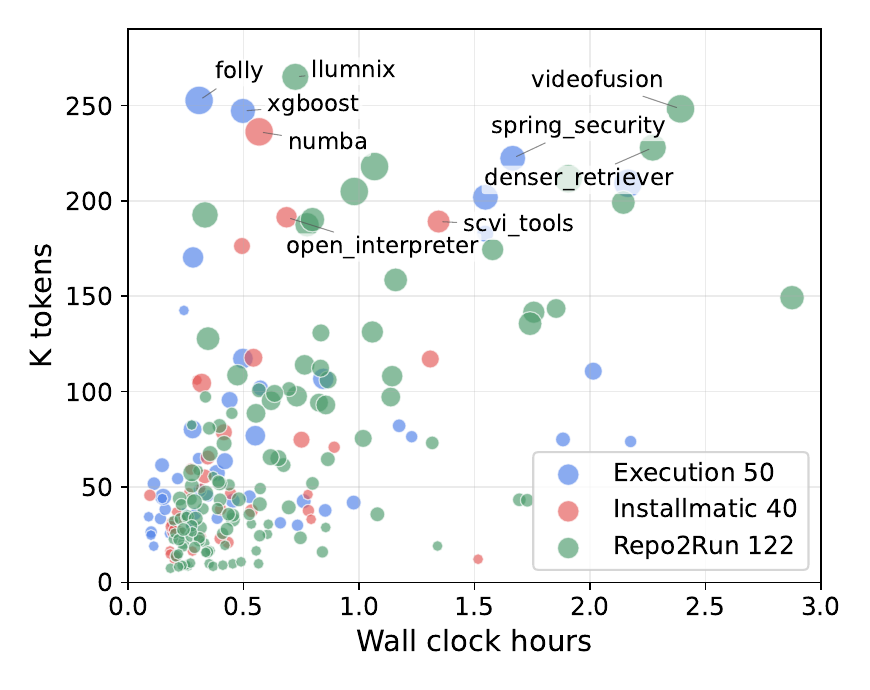}
  \caption{Relationship between token usage and time cost.}        
  \label{fig:token-time}
	\end{minipage}%
    \hfill
\end{figure*}

We further analyze the token and time cost of generating `.bootstrap` files and validating them with minimal verification.
Figure~\ref{fig:tokens} shows the distribution of total token usage per project, Figure~\ref{fig:wallclock} reports the corresponding time cost and Figure~\ref{fig:token-time} plots the relationship between token and time consumption.

\begin{table}[t]
\centering

\begin{minipage}[t]{0.48\linewidth}
  \centering
  \small
  \caption{Effectiveness across benchmarks.}
  \label{tab:effectiveness}
  \begin{tabular}{lrrr}
  \toprule
  \textbf{Benchmark} & \textbf{HerAgent} & \textbf{Ours} & \textbf{Delta} \\
  \midrule
  ExecutionAgent & 41/50 & 42/50 & +2.0\% \\
  Installamatic & 34/40 & 38/40 & +10.0\% \\
  Repo2Run & 104/122 & 117/122 & +10.7\% \\
    \midrule
   Overall & 179/212 & 197/212 & +8.5\% \\
  \bottomrule
  \end{tabular}
\end{minipage}
\hfill
\begin{minipage}[t]{0.48\linewidth}
  \centering
  \small
  \caption{Median time of cold/warm execution.}
  \label{tab:warm-cold-time}
  \begin{tabular}{lrrr}
  \toprule
  \textbf{Benchmark} &
  \textbf{Cold} &
  \textbf{Warm} &
  \textbf{Ratio} \\
  \midrule
  ExecutionAgent & 285.2s & 112.0s & 0.39$\times$ \\
  Installamatic & 337.8s & 282.3s & 0.84$\times$ \\
  Repo2Run & 391.9s & 309.8s & 0.79$\times$ \\
    \midrule
  Overall & 346.3s & 249.5s & 0.72$\times$ \\
  \bottomrule
  \end{tabular}
\end{minipage}

\end{table}

Across all 212 benchmark runs, BootstrapAgent uses a median of 43.8K tokens and 24.6 minutes per repository; the 90th percentile reaches 169.2K tokens and 79.0 minutes. The reported token usage is acceptable. Although BootstrapAgent introduces additional generation and validation overhead, this cost is largely paid once. Downstream agents can then reuse this artifact. This benefit is further amortized as the same repository is reused by more agents or users. The time cost includes model calls, network communication, package downloads, and minimal verification. Given that this time is comparable to the effort required for manual repository setup while producing a reusable build specification, we consider the overhead acceptable.

\subsection{Downstream bootstrap transfer}
\label{subsec:downstream-protocol}
We measure the impact of \texttt{.bootstrap} on downstream agents by comparing the cost of building projects with (warm) and without (cold) \texttt{.bootstrap}. We report two metrics: time and token usage. The detailed experiment setting is shown in Appendix~\ref{app:expset2}. As shown in Table~\ref{tab:warm-cold-time} and Table~\ref{tab:warm-cold-tokens}, warm-starting consistently reduces both time cost and token usage across all three benchmarks.

\begin{figure*}[t]
\centering

\begin{minipage}{0.32\textwidth}
  \centering
  \includegraphics[width=\linewidth]{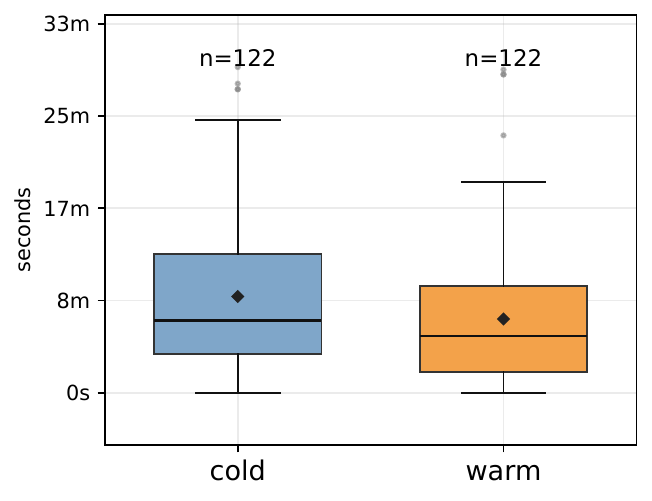}\\
  \small (a) \textsc{Repo2Run} wall time
\end{minipage}
\hfill
\begin{minipage}{0.32\textwidth}
  \centering
  \includegraphics[width=\linewidth]{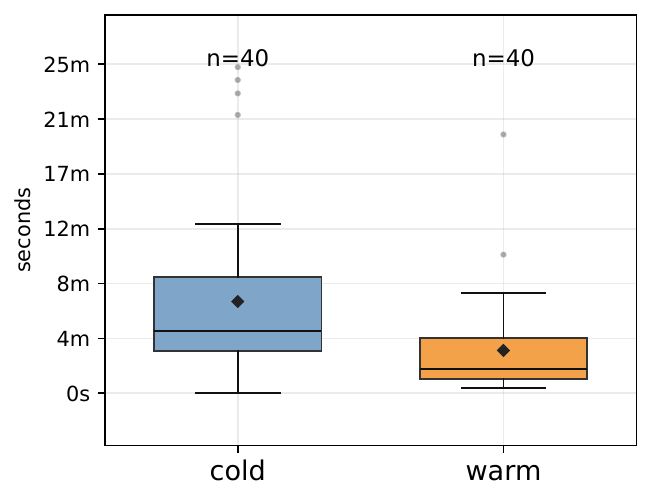}\\
  \small (b) \textsc{ExecutionAgent} wall time
\end{minipage}
\hfill
\begin{minipage}{0.32\textwidth}
  \centering
  \includegraphics[width=\linewidth]{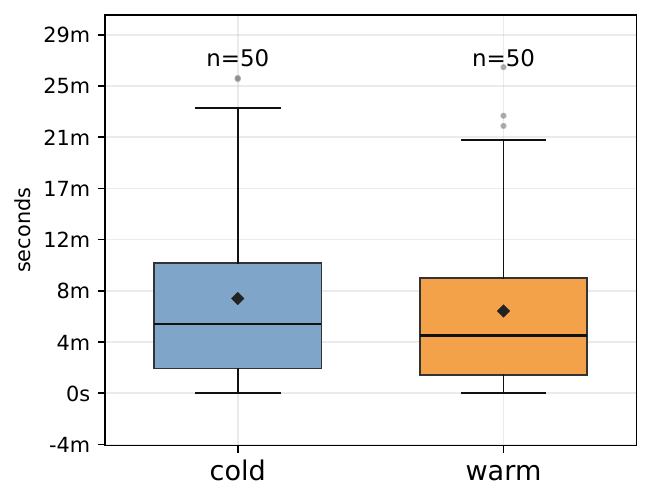}\\
  \small (c) \textsc{Installamatic} wall time
\end{minipage}


\begin{minipage}{0.32\textwidth}
  \centering
  \includegraphics[width=\linewidth]{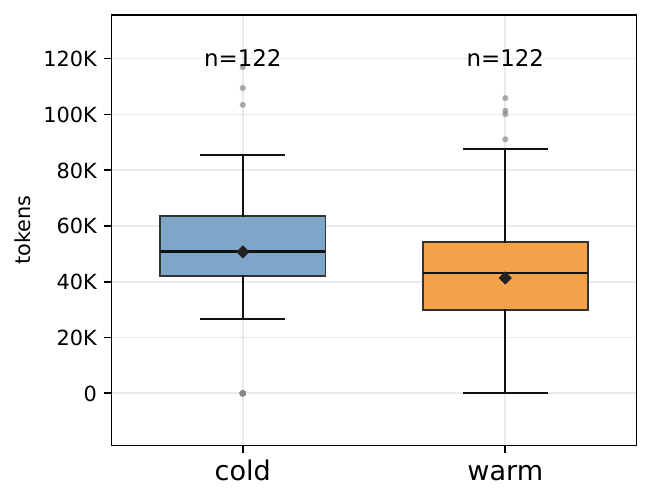}\\
  \small (d) \textsc{Repo2Run} active tokens
\end{minipage}
\hfill
\begin{minipage}{0.32\textwidth}
  \centering
  \includegraphics[width=\linewidth]{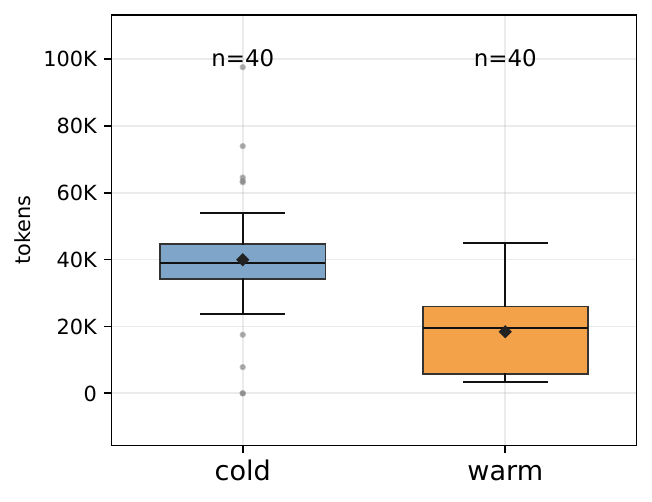}\\
  \small (e) \textsc{ExecutionAgent} active tokens
\end{minipage}
\hfill
\begin{minipage}{0.32\textwidth}
  \centering
  \includegraphics[width=\linewidth]{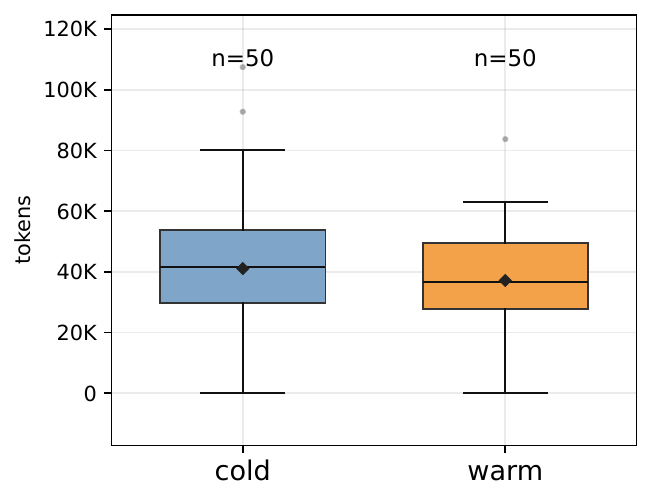}\\
  \small (f) \textsc{Installamatic} active tokens
\end{minipage}

\caption{Warm-starting results across benchmarks. The first row reports wall-clock time, and the second row reports active token usage. Warm starts consistently reduce resource consumption, with the largest gains on \textsc{ExecutionAgent}.}
\label{fig:warm-cold-all}
\end{figure*}

As shown in Figure~\ref{fig:warm-cold-all}, warm-starting consistently reduces downstream build-up cost across all three benchmarks. Warm-starting is an effective resource optimization for repository setup agents: median wall-time ratios range from 0.39$\times$ to 0.84$\times$, and median total-token ratios range from 0.19$\times$ to 0.66$\times$. The largest gain appears on \textsc{ExecutionAgent}, where median wall time drops from 285.2s to 112.0s (0.39$\times$), active tokens from 38.9K to 19.6K, and total tokens from 597.9K to 115.7K (0.19$\times$). \textsc{Installamatic} shows a smaller but consistent improvement, reducing median wall time from 337.8s to 282.3s (0.84$\times$), active tokens from 41.6K to 36.6K, and total tokens from 594.4K to 312.6K.
On \textsc{Repo2Run}, median time cost decreases from 391.9s to 309.8s (0.79$\times$), while median and total tokens drop from 50.8K to 43.1K and from 797.8K to 526.9K, respectively. 
\begin{table}[t]
\centering
\small
\caption{Token usage under cold and warm execution.}
\label{tab:warm-cold-tokens}
\begin{tabular}{lrrrrrr}
\toprule
\textbf{Benchmark} &
\textbf{Cold Median} &
\textbf{Warm Median} &
\textbf{Ratio} &
\textbf{Cold Total} &
\textbf{Warm Total} &
\textbf{Ratio} \\
\midrule
ExecutionAgent & 38.9K & 19.6K & 0.50$\times$ & 597.9K & 115.7K & 0.19$\times$ \\
Installamatic & 41.6K & 36.6K & 0.88$\times$ & 594.4K & 312.6K & 0.53$\times$ \\
Repo2Run & 50.8K & 43.1K & 0.85$\times$ & 797.8K & 526.9K & 0.66$\times$ \\
  \midrule
\textsc{Overall} & 46.3K & 36.9K & 0.80$\times$ & 710.7K & 419.4K & 0.59$\times$ \\
\bottomrule
\end{tabular}
\end{table}

In cold runs, the agent often has to rediscover the project layout, infer programming language, install dependencies, locate verification commands, and diagnose environment-specific failures. These steps create long tool-use and reasoning trajectories, increasing both time cost and token usage. Warm-starting reduces this repeated exploration by reusing prior setup state, cached observations, partially materialized environments, and existing build artifacts.

\subsection{Component-Targeted Ablation Study}
\label{subsec:ablation}

\newcommand{\drop}[1]{\textcolor{red}{\(\downarrow\) #1}}

\begin{table}[t]
\centering
\small
\caption{Component-targeted ablation of BootstrapAgent. Red numbers denote percentage-point drops relative to the full BootstrapAgent.}
\label{tab:ablation-components}
\setlength{\tabcolsep}{4pt}
\begin{tabular}{lccc}
\toprule
\textbf{Variant} &
\textbf{\textsc{Repo2Run}} &
\textbf{\textsc{ExecutionAgent}} &
\textbf{\textsc{Installamatic}} \\
\midrule
w/o Repository Discovery &
76.2\% \drop{19.7} &
52.0\% \drop{32.0} &
70.0\% \drop{25.0} \\
w/o Clean Replay &
95.1\% \drop{0.8} &
82.0\% \drop{2.0} &
90.0\% \drop{5.0} \\
w/o Trace-Driven Repair &
26.2\% \drop{69.7} &
18.0\% \drop{66.0} &
40.0\% \drop{55.0} \\
BootstrapAgent &
95.9\% &
84.0\% &
95.0\% \\
\bottomrule
\end{tabular}
\end{table}

\begin{figure*}[t]
  \centering
  \begin{minipage}[t]{0.32\linewidth}
      \centering
      \includegraphics[width=\linewidth]{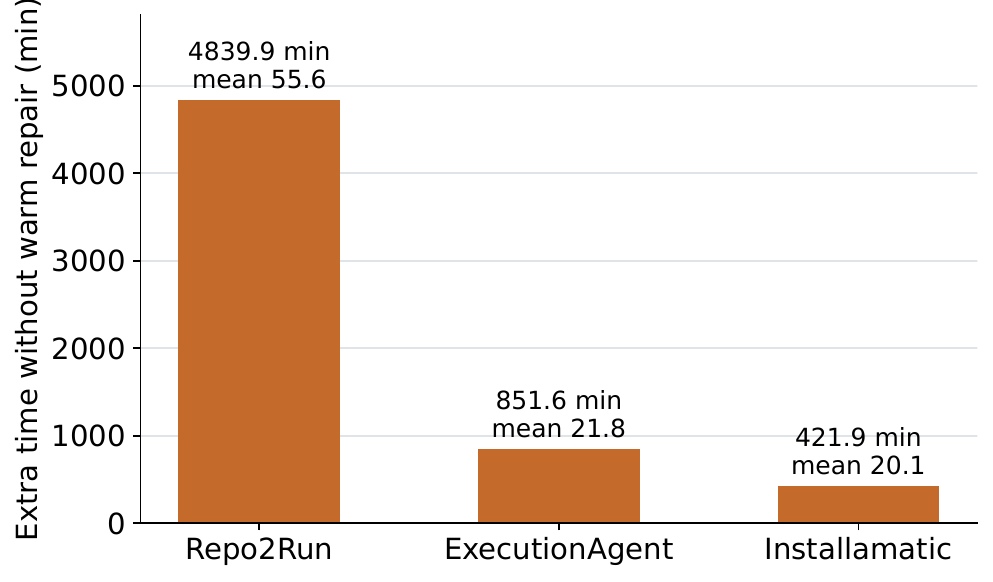}
      \caption{Estimated cost without warm repair.}
      \label{fig:ablation-warm-repair}
  \end{minipage}%
  \hfill
  \begin{minipage}[t]{0.32\linewidth}
      \centering
      \includegraphics[width=\linewidth]{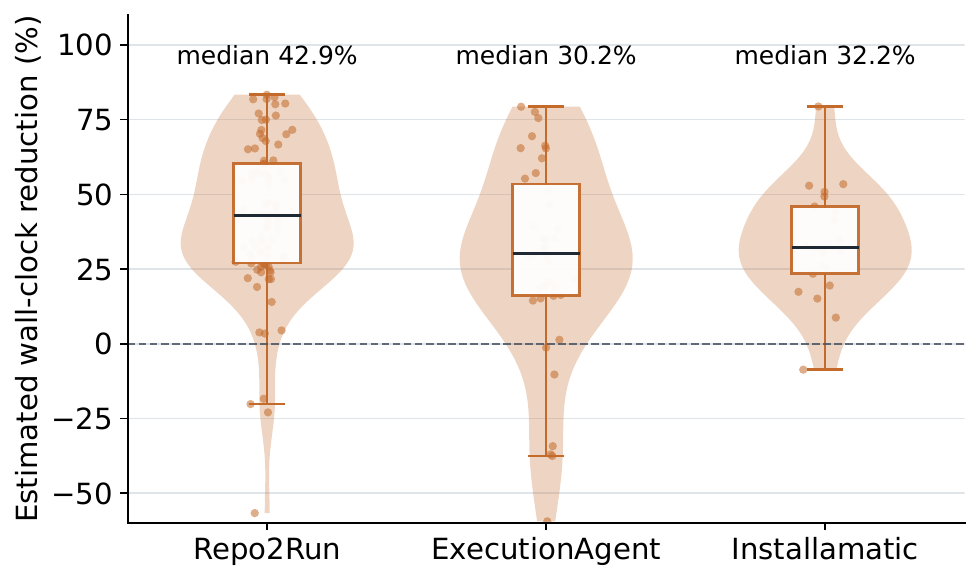}
      \caption{Distribution of warm-repair time reduction.}
      \label{fig:ablation-warm-repair2}
  \end{minipage}
  \hfill
  \begin{minipage}[t]{0.32\linewidth}
      \centering
      \includegraphics[width=\linewidth]{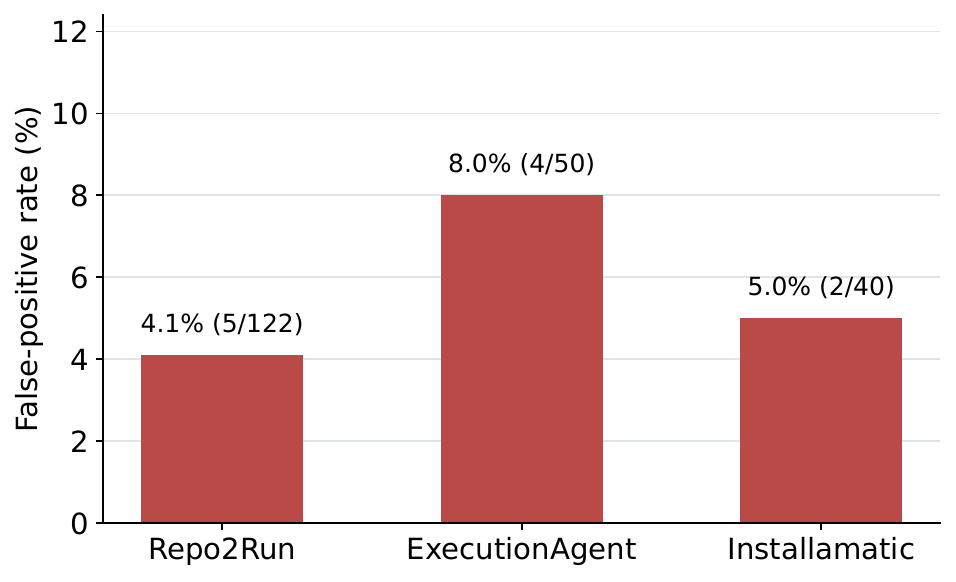}
      \caption{False positives caught by sanity check.}
      \label{fig:ablation-sanity-check}
  \end{minipage}
\end{figure*}
We conduct a component-targeted ablation study to isolate the contribution of mechanisms in BootstrapAgent: repository discovery, verification-guided repair, clean replay validation, warm-state repair, and sanity check.
For repository discovery, we ask the agent to generate the bootstrap plan directly, while keeping the rest of the pipeline unchanged.
For verification-guided repair, we disable iterative repair and evaluate only the first generated \texttt{.bootstrap} contract. For clean replay validation, we accept a successful warm repair directly without rerunning the final contract from scratch. For warm-state repair, each repair round is executed in a fresh Docker container instead of reusing the previous verified state. 
For sanity checks, we allow the repair agent to modify verification commands without checking the repaired contract against the original plan, evidence map, and strongest locally reproducible validation target.

We first analyze the impact of repository discovery, trace-driven repair, and clean replay on bootstrap success rate. As shown in Table~\ref{tab:ablation-components}, trace-driven repair has the largest impact: without iterative repair from execution traces, success drops by 69.7, 66.0, and 55.0 percentage points on the three benchmarks. This confirms that repository bootstrapping is an iterative trial-and-error process. Static repository evidence is often insufficient, and execution traces provide critical feedback for correcting missing dependencies, wrong working directories and incomplete verification scripts.

Repository discovery is also important. Removing the structured discovery report substantially reduces success, indicating that explicitly collecting bootstrap evidence before planning is more reliable than asking the agent to infer all setup information during generation. Clean replay has a smaller effect on the aggregate success rate, suggesting that warm-state repair usually preserves reproducibility. However, clean replay remains necessary because even a small number of warm-only successes may depend on cached packages, generated files, or residual container state, and should not be accepted as reusable bootstrap contracts.

We further analyze the time-saving effect of Warm-State Repair in Figures~\ref{fig:ablation-warm-repair} and~\ref{fig:ablation-warm-repair2}. Disabling Warm-State Repair would increase total time by 4839.9 minutes on \textsc{Repo2Run}, 851.6 minutes on \textsc{ExecutionAgent}, and 421.9 minutes on \textsc{Installamatic}. The per-repository distribution further shows median wall-clock reductions of 42.9\%, 30.2\%, and 32.2\% on the three benchmarks, respectively. Overall, Warm-State Repair does not mainly improve success rate; instead, it makes trace-driven iteration computationally practical by avoiding repeated full setup costs.

Beyond success rate and efficiency, we examine whether the accepted \texttt{.bootstrap} artifacts remain trustworthy. Figure~\ref{fig:ablation-sanity-check} shows that sanity checks catch 5/122 such cases on \textsc{Repo2Run}, 4/50 on \textsc{ExecutionAgent}, and 2/40 on \textsc{Installamatic}. This mechanism improves reliability not by increasing apparent success, but by preventing reward-hacked or drifted repairs from being accepted as valid bootstrap contracts.

Overall, trace-driven repair provides the main success-rate gain, repository discovery improves the quality of the initial plan, and warm-state repair reduces the cost of repeated verification. Clean replay and sanity checks contribute primarily to trustworthiness: clean replay prevents warm-state artifacts from being mistaken for reproducible contracts, while sanity checks prevent repair drift and verification weakening. These results show that BootstrapAgent relies on both adaptive repair for effectiveness and deterministic safeguards for reproducible, trustworthy bootstrap generation.

\section{Discussion}
\label{sec:discussion}

\subsection{Failure Analysis}
\label{subsec:discussion-failure-analysis}

Repositories may either omit essential build information, such as the required Python version, or provide inconsistent setup information across files. Both cases can lead the agent to choose the wrong environment at the start, and later repairs often fail because the resulting errors stem from version incompatibility rather than simple command mistakes. These failures explain why BootstrapAgent uses conservative acceptance criteria. On large or underspecified repositories, repeated repair can produce brittle fixes that silence the current error without yielding a transferable bootstrap, such as skipping a dependency, guessing an arbitrary pin, weakening verification, or relying on warm-container state. Clean replay and strongest-verification guardrails may lower the apparent solve rate, but they prevent overfitted \texttt{.bootstrap} files from being accepted as reusable startup
knowledge.

\subsection{Limitations}
\label{subsec:limitations}

BootstrapAgent establishes local bootstrap readiness rather than full CI reproduction or semantic correctness. This boundary is unavoidable for a broad benchmark: many real CI pipelines require secrets, external services or long-running jobs that cannot be made uniformly available without changing the task from repository bootstrapping to full system reproduction. Additionally, once a .bootstrap file is generated, it becomes part of the repository documentation. Like other documentation, it must be maintained as the project evolves; otherwise, it may become outdated and mislead future agent bootstrapping or manual setup.

\subsection{Broader Impacts}
\label{subsec:broader-impacts}

BootstrapAgent can reduce duplicated setup effort, failed commands, token use, and wasted compute when developers or coding agents work with unfamiliar repositories, especially public projects where users repeatedly rediscover the same setup and verification details. The main risk is over-trusting generated bootstrap contracts, which may encode transient workarounds, outdated dependencies, network assumptions, or weak validation. BootstrapAgent mitigates this through deterministic Docker verification, clean replay, strong-verification guardrails, structured traces, and warnings for fragile commands. Maintainer review remains necessary before adopting generated setup instructions in production documentation.

\section{Conclusion}
\label{sec:conclusion}

We introduced \textbf{BootstrapAgent}, a multi-agent framework that turns repository bootstrapping from an ephemeral trial-and-error interaction into reusable startup knowledge. Instead of requiring each future agent to rediscover runtime versions, dependency constraints, working directories, diagnostic checks, and verification commands, BootstrapAgent distills these facts into a persistent and verifiable \texttt{.bootstrap} contract. Through repository discovery, structured planning, deterministic Docker verification, trace-driven repair, warm repair with clean replay, and two-level verification, the generated contract helps downstream agents save token and time cost when building environment. Our evaluation on 212 repositories shows that this framework is effective in practice. These results suggest that repository setup should be treated not merely as a one-time environment construction task, but as an amortizable form of agent knowledge. At the same time, a successful \texttt{.bootstrap} contract does not replace full CI, maintainer judgment, or complete functional validation. Future work can extend this direction with stronger dependency solving, historical environment inference, external artifact recovery, and richer support for repositories requiring services, GPUs, or large-scale integration tests.

\clearpage
\bibliographystyle{plainnat}
\bibliography{references}

\newpage
\appendix

\section{Key Prompts}
\label{app:key-prompts}

This appendix lists the key prompts used by our agentic bootstrap-generation pipeline.
The prompts are shown in template form: repository-specific discovery reports, CI evidence,
current plans, and verifier traces are inserted at runtime.
Together, these prompts cover repository evidence collection, CI-derived validation extraction,
structured command planning, verifier-guided repair, two-level verification, and
language-specific setup guardrails.

\begin{PromptBox}{Main Bootstrap Agent System Prompt}
You are the MainBootstrapAgent.

Coordinate specialized subagents to generate a verified .bootstrap package for
an unfamiliar repository. All machine-consumed outputs must be valid JSON or
YAML matching the provided schemas. Do not modify business source code. The
deterministic verifier is the only authority on success or failure.
\end{PromptBox}

\begin{PromptBox}{Evidence Collection Prompts}
Discovery agent:
Collect repository evidence only. Return DiscoveryReport JSON.

CI evidence agent:
Inspect CI files and return CIEvidenceReport YAML.

Planner agent:
Generate BootstrapPlan JSON from discovery and CI evidence.

Repair agent:
Generate RepairPlan JSON from a failed verifier trace.
\end{PromptBox}

\begin{PromptBox}{Command Planner System Prompt}
You are CommandPlannerAgent. Return only the structured BootstrapPlan.

{COMMAND_CONSTRAINTS}
\end{PromptBox}

\begin{PromptBox}{Command Planner User Prompt}
Generate a BootstrapPlan JSON for this repository. Use only commands that are plausible
from the provided evidence. Preserve provenance in source/reason fields. The plan must
describe bootstrap commands for the checked-out source. Commands that mutate the checkout
or use remote installers are allowed when necessary and will be logged as safety warnings.

{COMMAND_CONSTRAINTS}

DiscoveryReport:
{DISCOVERY_REPORT_JSON}

CIEvidenceReport:
{CI_EVIDENCE_REPORT_JSON}
\end{PromptBox}

\begin{PromptBox}{Repair Agent System Prompt}
You are RepairAgent. Return only the structured RepairPlanDelta. Do not call tools.

{COMMAND_CONSTRAINTS}
\end{PromptBox}

\begin{PromptBox}{Repair Agent User Prompt}
Generate a RepairPlanDelta JSON from this failed verifier result. Return only the smallest
delta needed to repair the current BootstrapPlan; omitted or null fields are copied from the
current plan locally. Prefer replace_commands for one-command edits, move_commands for reordering,
and insert_commands or remove_commands for list changes. Use replace_doctor or replace_install only
when the whole list must change. For doctor/install list edits, use the zero-based index values shown
in the compact JSON and never remove an index that is not present. Only change setup, verify, doctor
commands, cwd, timeout, agent_context, evidence, or failure_playbook. Checkout mutations or remote
installers are allowed when needed for the repository bootstrap and will be logged as safety warnings.
Do not call tools or inspect files; use only the supplied JSON. Focus on the failed command and the
shortest repair.

Current BootstrapPlan compact JSON:
{CURRENT_BOOTSTRAP_PLAN_JSON}

VerifierResult compact JSON:
{VERIFIER_RESULT_JSON}
\end{PromptBox}

\begin{PromptBox}{Core Command Constraints}
Command constraints:
- Assume the verifier starts from a fresh, minimal Ubuntu container. Do not assume project users
  already have language runtimes, compilers, package managers, build tools, or test tools installed.
  If setup or verify needs a tool, install or enable it in install commands before first use.
- Doctor commands run after setup.sh and should normally be read-only health checks of the prepared
  environment. Put environment setup, package installation, dependency installation, and build steps
  only in install commands.
- Every command reason must match the actual command string.
- Put validation, import, and test commands only in minimal_verify, strongest_verify, or run_probe.
- minimal_verify should be the lowest-cost trustworthy project check and is a hard verification gate.
  strongest_verify is advisory and should be the strongest reproducible local CI-derived validation.
- Prefer the repository's native development workflow for a fresh source checkout.
- Do not install a published package with the same name as the repository as a substitute for checking
  the checked-out source.
- Do not use a runtime version check, such as python3 --version or node --version, as verification for
  a non-empty source repository.
- The verifier runs setup.sh, doctor.sh, and verify.sh sequentially in the same Docker container.
  Environment changes from setup.sh are available to doctor.sh and verify.sh.
- Prefer cwd "."; do not use host paths in commands.
- Commands that look risky are executed and logged in .bootstrap/safety_warnings.json rather than
  rejected by policy.
\end{PromptBox}

\begin{PromptBox}{Language-Specific Guardrails}
Additional project profiles are appended when repository evidence indicates Bazel/C/C++, Node.js,
Java, Rust, Go, or native C/C++ workflows. These profiles require the agent to use the checked-out
source workflow, install missing toolchains before use, prefer lockfile-indicated package managers,
and avoid replacing project validation with unrelated runtime version checks.
\end{PromptBox}

\section{Benchmark Collections}
\label{app:benchmark}
We evaluate BootstrapAgent on three public benchmarks that cover complementary repository setup scenarios. Table~\ref{tab:benchmark-overview} summarizes their sizes and language distributions.

\begin{table}[t]
\centering
\caption{Benchmark overview. ExecutionAgent-Bench is reported by main language in our benchmark metadata; the original paper notes that the 50 projects contain 14 programming languages in total.}
\label{tab:benchmark-overview}
\begin{tabular}{llp{6.2cm}}
\toprule
Benchmark & Size & Language Distribution \\
\midrule
Repo2Run-Bench & 122 & Python: 122 \; (100\%) \\
ExecutionAgent-Bench & 50 & JavaScript: 12 \; (24\%), Python: 11 \; (22\%), C: 10 \; (20\%), Java: 9 \; (18\%), C++: 8 \; (16\%) \\
Installamatic-Bench & 40 & Python: 40 \; (100\%) \\
\bottomrule
\end{tabular}
\end{table}

Repo2Run-Bench~\citep{hu2025repo2run} is a Python benchmark centered on pytest-oriented environment construction. The original Repo2Run dataset contains 420 Python repositories; following the cross-benchmark setting used by recent work, we evaluate on the 122-repository subset used for comparison. This benchmark stresses Python dependency installation and test-command discovery.

ExecutionAgent-Bench~\cite{test} contains 50 repositories with diverse build and test ecosystems. In our benchmark metadata, the repositories are grouped into five main-language categories: JavaScript, Python, C, Java, and C++. The original benchmark further reports that these projects contain 14 programming languages when considering the full repository and test code. We use this benchmark to evaluate whether BootstrapAgent generalizes beyond Python-only setup.

Installamatic-Bench ~\citep{installmatic} contains 40 Python repositories with manually studied installation procedures and test suites. This benchmark is closely aligned with our target setting because it focuses on repository installation and validation rather than downstream patch generation.

\section{Experiment Setting}
\label{app:expset}

\subsection{Effectiveness Evaluation}
\label{app:expset1}

We evaluate the effectiveness of BootstrapAgent under a fixed execution budget and a controlled verifier environment. For each repository, the system is allowed to iteratively repair the generated bootstrap package until either verification succeeds or the configured budget is exhausted. The main budget parameters are shown in Table~\ref{tab:exp-config}.

\begin{table}[t]
\centering
\small
\caption{Configuration used in the effectiveness evaluation.}
\label{tab:exp-config}
\begin{tabular}{l r}
\toprule
Parameter & Value \\
\midrule
Maximum repair loops & 20 \\
Maximum clean-replay repair loops & 3 \\
Maximum strongest-test repair attempts & 5 \\
Maximum structured LLM retries & 5 \\
Maximum repair structured LLM retries & 5 \\
Initial LLM timeout & 300 s \\
Repair LLM timeout & 180 s \\
Maximum shell commands & 80 \\
Maximum total wall-clock time & 3600 s \\
Doctor command timeout & 120 s \\
Setup command timeout & 3600 s \\
Minimal verification timeout & 300 s \\
timeout & 1200 s \\
\bottomrule
\end{tabular}
\end{table}

All verification runs are executed inside a deterministic Docker environment based on Ubuntu 24.04. The
image contains only common system utilities required for repository inspection, dependency installation,
and command execution, including \texttt{bash}, \texttt{curl}, \texttt{git}, \texttt{make}, \texttt{tar},
\texttt{unzip}, and related core Unix tools. The working directory inside the container is fixed to
\texttt{/workspace/repo}. This minimal environment reduces interference from host-specific packages and
ensures that successful verification depends on the generated bootstrap instructions rather than on pre-
existing local state.

\subsection{Downstream Agent Evaluation}
\label{app:expset2}

To evaluate whether BootstrapAgent helps downstream coding agents, we conduct a paired controlled experiment
using Claude Code as a representative code agent. Claude Code is executed in headless mode with session
persistence disabled, so that previous conversations or historical context do not affect the results.

For each project, we compare two conditions. In the \emph{warm} condition, the project directory contains
the pre-generated \texttt{.bootstrap/} directory. The agent is instructed to read \texttt{.bootstrap/
commands.json} and directly execute the provided \texttt{minimal\_verify} and \texttt{strongest\_verify}
commands. In the \emph{cold} condition, the original \texttt{.bootstrap/} directory is temporarily hidden.
The agent must identify the programming language, build system, dependency requirements, and environment
setup procedure from scratch. However, it is still given the same final verification commands from the
original \texttt{commands.json}, ensuring that the warm and cold settings share the same target task.

Each run uses the same timeout and budget limits, namely 1800 seconds and a maximum API budget of 3 USD.
After each cold run, the original \texttt{.bootstrap/} directory is restored to avoid contamination of
subsequent experiments by files generated during the cold condition.

We record the structured JSON output produced by Claude Code, including total elapsed time, API time,
input tokens, output tokens, cache-read tokens, number of interaction turns, and estimated cost. In
addition, the agent is asked to report the final task status, the number of failed commands, and the
number of search-related commands. We quantify the benefit of BootstrapAgent by comparing the paired cold and
warm runs for the same project. For example, time savings are computed as
\[
T_{\text{cold}} - T_{\text{warm}},
\]
token savings as
\[
C_{\text{cold}} - C_{\text{warm}},
\]
and cost savings as
\[
P_{\text{cold}} - P_{\text{warm}}.
\]
Because every project is evaluated once under each condition, this forms a within-project paired
comparison, reducing confounding effects from differences in project size, language ecosystem, dependency
complexity, and test-suite cost.
\normalfont\normalsize\rmfamily\normalcolor
\section{Additional Reproducibility and Ethics Details}
\label{app:discussion-details}

\subsection{Reproducibility Details}
\label{app:reproducibility-details}
BootstrapAgent is evaluated as a repository bootstrapping system rather than as a model training method. The input to each run is a public GitHub repository from one of three external benchmark collections: Repo2Run-Bench, ExecutionAgent-Bench, and Installamatic-Bench. The output is a repository-local \texttt{.bootstrap} contract plus an evaluation log. The contract contains setup, diagnostic, and verification commands; the log records success or failure, stage outcomes, retry counts, token usage, time cost, and verifier traces.

All agentic planning and repair runs use \texttt{DeepSeek-V4-Flash-2026-04-25} with temperature 1.0. The implementation exposes the same setting through the \texttt{rethink bootstrap} and \texttt{rethink batch} commands, with \texttt{deepseek:deepseek-chat} used as the provider identifier in the released scripts. The default budgets are 20 repair rounds, 3 clean-replay repair rounds, 5 strongest-verification repair attempts, 5 structured-output retries for initial plans, 5 structured-output retries for repairs, 80 shell commands, and a 3600 second total time budget. LLM requests use a 300 second timeout for initial planning and a 180 second timeout for repair planning.

Verification is performed inside a Docker container built from an Ubuntu 24.04 base image. The base image includes only general-purpose development utilities such as \texttt{bash}, \texttt{git}, \texttt{curl}, \texttt{wget}, \texttt{make}, archive tools, and standard Unix file utilities. Language runtimes, compilers, package managers, project dependencies, and test tools are not assumed to be preinstalled; the generated \texttt{.bootstrap/setup.sh} must provision them when needed. Stage timeouts are 3600 seconds for setup, 120 seconds for doctor commands, 300 seconds for minimal verification, and 1200 seconds for strongest verification.

Success requires clean replay: \texttt{setup}, \texttt{doctor}, and \texttt{minimal\_verify} must pass from a fresh container. \texttt{strongest\_verify} is recorded when available and used as a guardrail against validation downgrades, but it is not required as the universal success gate because many real CI workflows depend on secrets, service containers, specialized hardware, or long-running jobs. External services, private credentials, GPU requirements, and expensive training workloads are outside the default success criterion.

\subsection{External Asset Use and Redistribution}
\label{app:benchmark-assets}
The benchmark composition is documented in Appendix~\ref{app:benchmark}. The experiments use those existing public repository benchmarks rather than a newly constructed private benchmark.

The released research artifacts include the BootstrapAgent prototype, run scripts, benchmark URL lists, generated \texttt{.bootstrap} contracts, and summarized evaluation logs. We do not redistribute third-party repository source code as part of the paper artifact. Each upstream repository remains owned by its original maintainers and governed by its original license and terms of use. The URL lists are intended to let researchers recover the same public repositories through the upstream benchmark sources or GitHub, subject to those upstream terms.

\subsection{Data, Ethics, and Asset Licenses}
\label{subsec:data-ethics}
The research uses public GitHub software artifacts from existing benchmark collections. It does not involve private data, participant recruitment, crowdsourcing, surveys, or intervention with human subjects. BootstrapAgent may inspect repository files and public CI configuration to infer setup and verification commands, but the released paper artifact does not repackage third-party source code.

Existing repositories are credited by URL in the benchmark lists and remain governed by their original licenses and terms. The new assets introduced by this work are the BootstrapAgent prototype, the \texttt{.bootstrap} contract format, run scripts, generated contracts, and evaluation summaries. These artifacts are separate from the upstream repositories used as evaluation inputs.

\subsection{LLM Usage}
\label{app:llm-usage}
LLMs are a core component of the method and evaluation. BootstrapAgent uses LLM-based agents for repository evidence interpretation, command planning, and trace-driven repair. The downstream transfer experiments also use coding agents to compare cold repository setup against setup with a generated \texttt{.bootstrap} contract. LLMs are therefore part of the scientific method under evaluation, not merely a writing or formatting aid.


\end{document}